\documentclass[a4paper,10pt,eqno]{article}
\usepackage{theorem}
\usepackage{latexsym,amssymb,amsfonts,amsmath}
\theoremheaderfont{\scshape}

\newcommand{\C}{{\mathbb C}}

\newcommand{\xvec}{\ensuremath{\boldsymbol{x}}}

\newcommand{\evec}{\ensuremath{\boldsymbol{e}}}

\def\x{\mbox{\boldmath$x$}}

\def\e{\mbox{\boldmath$e$}}
\newcommand{\Z}{{\mathbb Z}}
\def\zero{\mbox{\boldmath$0$}}

\numberwithin{equation}{section}

\begin{document}
\title{{\Large {\bf On the support of the Grover walk on higher-dimensional lattices}}}
\author{
{\small Norio Konno\footnote{konno-norio-bt@ynu.ac.jp},\quad Sarato Takahashi\footnote{takahashi-sarato-vb@ynu.jp(e-mail of the corresponding author)}}\\
{\scriptsize  Department of Applied Mathematics, Faculty of Engineering, Yokohama National University}\\
{\scriptsize \footnotesize\it 79-5 Tokiwadai, Hodogaya, Yokohama, 240-8501, Japan}\\
}
\date{}
\maketitle
\par\noindent
\par\noindent
{\bf Abstract}. This paper presents the minimum supports of states for stationary measures of the Grover walk on the $d$-dimensional lattice by solving the corresponding eigenvalue problem. The numbers of the minimum supports for moving and flip-flop shifts are $2^d \ (d\ge1)$ and $4 \ (d\ge2)$, respectively.
 



\section{Introduction}

%
%
%
%
%

The quantum walk was introduced by Aharonov et al. \cite{Aharonov1} as a generalization of the random walk on graphs. On the one-dimensional lattice $\Z$, where $\Z$ is the set of integers, the properties of quantum walks are well studied, see Konno \cite{Konno}, for example. There are some results on the Grover walk on $\Z^2$, such as weak limit theorem by Watabe et al. \cite{Watabe et al.} (moving shift case) and Higuchi et al. \cite{Higuchi et al.} (flip-flop shift case), and localization shown by Inui et al. \cite{Inui et al.} (moving shift case) and Higuchi et al. \cite{Higuchi et al.} (flip-flop shift case).

In this paper, we present the minimum support of states for the stationary measures of the Grover walk on $\Z^d$ by solving the corresponding eigenvlaue problem. As for the number of the support of the Grover walk on $\Z^d$ with moving shift, $2^2$ ($\Z^2$ case) and $3^d$ ($\Z^d$ case with $d\ge2$) were given in Stefanak et al. \cite{Stefanak1} and Komatsu and Konno \cite{Komatsu and Konno.} by the Fourier analysis, respectively. Compared with the above-mentioned previous results, the number of our minimum support for $\Z^d$ case with $d\ge1$ is $2^d$ (Theorem 1). Moreover, concerning the number of the support of the Grover walk on $\Z^d$($d\ge2$) with flip-flop shift, 4 was obtained in Higuchi et al. \cite{Higuchi et al.} by the spectral mapping theorem, which coincides with our result (Theorem 2). Remark that any finite support does not exist for $\Z^1$ case. 

The rest of the paper is as follows. Section 2 is devoted to the definition of the discrete-time quantum walk on $\Z^d$. Section 3 deals with the stationary measure of the Grover walk on $\Z^d$. We give main results on minimum support for the Grover walk on $\Z^d$ with moving shift (Theorem 1) in Section 4 and flip-flop shift (Theorem 2) in Section 5, respectively. Section 6 summarizes our paper..

\section{Discrete-time quantum walks on $\mathbb{Z}^d$}\label{quantumzddef}
In this section, we give the definition of $2d$-state discrete-time quantum walks on $\mathbb{Z}^d$. 
The quantum walk is defined by using a shift operator and a unitary matrix. Let $\mathbb{C}$ be the set of complex numbers. For $i\in\{1,2,\ldots,d\}$, 
the shift operator $\tau_i$ is given by 
\begin{align}
(\tau_if)(\xvec)=f(\xvec-\evec_{i})\quad (f:\mathbb{Z}^d\longrightarrow\mathbb{C}^{2d}, \ \xvec\in\mathbb{Z}^d),
\label{eq2.1}
\end{align} 
where $\{\evec_1,\evec_2,\ldots,\evec_d\}$ denotes the standard basis of $\mathbb{Z}^d$. Let $A=(a_{ij})_{i,j=1,2,\ldots,2d}$ be a $2d\times 2d$ unitary matrix. 
We call this unitary matrix the coin matrix. To describe the time evolution of the quantum walk, decompose the unitary matrix $A$ as
\begin{align}
A=\sum_{i=1}^{2d}P_{i}A,
\label{eq2.2}
\end{align}
where $P_i$ denotes an orthogonal projection onto the one-dimensional subspace $\mathbb{C}\eta_i$ in $\mathbb{C}^{2d}$. Here $\{\eta_1,\eta_2,\ldots,\eta_{2d}\}$ denotes the standard basis on $\mathbb{C}^{2d}$. 
The walk associated with the coin matrix $A$ is given by
\begin{equation}
U_A=\sum_{i=1}^d\Big(P_{2i-1}A\tau_{i}^{-1}+P_{2i}A\tau_{i}\Big).
\label{eq2.3}
\end{equation}

Let $\Z_{\ge}=\{0,1,2,\ldots\}$. The state at time $n\in\Z_{\ge}$ and location $x\in\Z^d$ can be expressed by a $2d$-dimensional vector:
\begin{align}
\Psi_{n}(\xvec)={}^{T} \begin{bmatrix}\Psi^{1}_{n}(\xvec), \Psi^{2}_{n}(\xvec), \cdots, \Psi^{2d}_{n}(\xvec) \end{bmatrix} \in\mathbb{C}^{2d} ,
\label{eq2.4}
\end{align} 
where $T$ denotes a transposed operator. For $\Psi_n:\mathbb{Z}^d\longrightarrow\mathbb{C}^{2d} \ (n\in\mathbb{Z}_{\geq})$, it follows from Eq. \eqref{eq2.3} that
\begin{align}
\Psi_{n+1}(\xvec)\equiv(U_A\Psi_{n})(\xvec)=\sum_{i=1}^{d}\Big(P_{2i-1}A\Psi_{n}(\xvec+\evec_i)+P_{2i}A\Psi_{n}(\xvec-\evec_i)\Big).
\label{eq2.5}
\end{align} 
This equation means that the particle moves at each step one unit to the $x_i$-axis direction with matrix $P_{2i}A$ or one unit to the $-x_i$-axis direction with matrix $P_{2i-1}A$. For time $n\in\mathbb{Z}_{\geq}$ and location $x\in\mathbb{Z}^d$, we define the measure $\mu_n(\xvec)$ by	
\begin{align}
\mu_n(\xvec)=\|\Psi_n(\xvec)\|_{\mathbb{C}^{2d}}^2,
\label{eq2.6}
\end{align}
where $\|\cdot\|_{\mathbb{C}^{2d}}$ denotes the standard norm on $\mathbb{C}^{2d}$. Let $\mathbb{R}_{\geq}=[0,\infty)$.  Here we introduce a map $\phi:(\mathbb{C}^{2d})^{\mathbb{Z}^d}\longrightarrow (\mathbb{R}_{\geq})^{\mathbb{Z}^d}$ such that if $\Psi_n:\mathbb{Z}^d\longrightarrow\mathbb{C}^{2d}$ and $\xvec\in\mathbb{Z}^{d}$, 
thus we get
\begin{align}
\phi(\Psi_n)(\xvec)=\sum_{j=1}^{2d}|\Psi_n^{j}(\xvec)|^2=\mu_n(\xvec),
\label{eq2.7}
\end{align}
namely this map $\phi$ has a role to transform from amplitudes to measures.

\section{Stationary measure of the Grover walk on $\mathbb{Z}^d$}\label{zdstationary1}
\noindent In this section, we give the definition of the stationary measure for the quantum walk. We define a set of measures, $\mathcal{M}_s(U_A)$, by
\begin{equation}
\mathcal{M}_s(U_A)=\Big\{\mu\in[0,\infty)^{\mathbb{Z}^d}\setminus\{\textbf{0}\};\ there\ exists\ \Psi_0\in\left(\mathbb{C}^{2d}\right)^{\mathbb{Z}^d}\ such\ that \ 
\phi(U_A^n\Psi_0)=\mu \ \  (n\in \Z_{\ge})\Big\},
\label{eq3.1}
\end{equation}
where $\textbf{0}$ is the zero vector. Here $U_A$ is the time evolution operator of quantum walk associated with a unitary matrix $A$. We call this measure $\mu\in\mathcal{M}_s(U_A)$ the stationary measure for the quantum walk defined by the unitary operator $U_A$. If $\mu\in\mathcal{M}_s(U_A)$, then $\mu_n=\mu$ for $n\in\mathbb{Z}_{\geq}$, where $\mu_n$ is the measure of quantum walk given by $U_A$ at time $n$.

Next we consider the following eigenvalue problem of the quantum walk determined by $U_A$:
 \begin{equation}
U_A\Psi=\lambda\Psi\quad(\lambda\in\mathbb{C},\ |\lambda|=1).
\label{eq3.2}
\end{equation}
We introduce the set of solutions of Eq. \eqref{eq3.2} for $\lambda\in\C$ with $|\lambda|=1$ as follows.
\begin{equation}
W(\lambda)=\{\Psi\ne\zero:U_A\Psi=\lambda\Psi\}.
\label{eq3.3}
\end{equation}
Then for $\Psi\in W(\lambda)$, we see that $\phi(\Psi)\in\mathcal{M}_s(U_A)$. If the function $\Psi$ satisfied with $\lambda=1$ in Eq. \eqref{eq3.2}, then  $\Psi$ is called the {\it stationary amplitude}. From now on, we focus on the Grover Walk on $\Z^d$ which is defined by the following $2d \times 2d$ coin matrix $U_G=(g_{ij})_{i,j=1,2,\ldots,2d}$ with 
\begin{align}
g_{ij}=\frac{1}{d}-\delta_{ij}.
\label{eq3.4}
\end{align}
Remark that Komatsu and Tate \cite{Komatsu and Tate.} showed that the eigenvalue of Eq. \eqref{eq3.2} is only $\lambda=\pm1$ for the $d$-dimensional Gorver walk. Our purpose of this paper is to investigate the support of the $2d$-state Grover walk on $\Z^d$.

\section{Grover walk on $\Z^d$ with moving shift}

In this section, we present our main results on the support of the Grover walk on $\Z^d$ with moving shift. To do so, we begin with the eigenvalue problem $U_{G}\Psi=\lambda\Psi \ (\lambda \in \C \ {\rm with} \ |\lambda|=1)$, which is equivalent to 

\begin{equation}
\left\{
\begin{gathered}
\lambda\Psi^{1}(\xvec)
=\frac{1-d}{d}\Psi^{1}(\xvec+\evec_1)+\frac{1}{d}\Psi^{2}(\xvec+\evec_1)
+\cdots
+\frac{1}{d}\Psi^{2d-1}(\xvec+\evec_1)+\frac{1}{d}\Psi^{2d}(\xvec+\evec_1),\\
\lambda\Psi^{2}(\xvec)
=\frac{1}{d}\Psi^{1}(\xvec-\evec_1)+\frac{1-d}{d}\Psi^{2}(\xvec-\evec_1)
+\cdots
+\frac{1}{d}\Psi^{2d-1}(\xvec-\evec_1)+\frac{1}{d}\Psi^{2d}(\xvec-\evec_1),\\
\vdots\\
\lambda\Psi^{2d-1}(\xvec)
=\frac{1}{d}\Psi^{1}(\xvec+\evec_d)+\frac{1}{d}\Psi^{2}(\xvec+\evec_d)
+\cdots
+\frac{1-d}{d}\Psi^{2d-1}(\xvec+\evec_d)+\frac{1}{d}\Psi^{2d}(\xvec+\evec_d),\\
\lambda\Psi^{2d}(\xvec)
=\frac{1}{d}\Psi^{1}(\xvec-\evec_d)+\frac{1}{d}\Psi^{2}(\xvec-\evec_d)
+\cdots
+\frac{1}{d}\Psi^{2d-1}(\xvec-\evec_d)+\frac{1-d}{d}\Psi^{2d}(\xvec-\evec_d),\\
\end{gathered}
\right.
\label{eq4.1}
\end{equation}
where $\Psi(\xvec)={}^{T} \begin{bmatrix}\Psi^{1}(\xvec), \Psi^{2}(\xvec), \cdots, \Psi^{2d}(\xvec)\end{bmatrix}$ $(\x \in \Z^d)$. Put ${\displaystyle \Gamma(\x)= \sum_{j=1}^{2d} \Psi^{j} (\x)}$ for $\x\in\Z^d$. By using $\Gamma(\x)$, Eq. (\ref{eq4.1}) can be written as

\begin{equation}
\begin{aligned}
&\lambda\Psi^{2k-1}(\xvec-\evec_k)+\Psi^{2k-1}(\xvec)=\frac{1}{d}\Gamma(\xvec),\\
\end{aligned}
\label{eq4.2}
\end{equation}
\begin{equation}
\begin{aligned}
&\lambda\Psi^{2k}(\xvec+\evec_k)+\Psi^{2k}(\xvec)=\frac{1}{d}\Gamma(\xvec),
\end{aligned}
\label{eq4.3}
\end{equation}
for any $k=1,2,\dots,d$ and $\x \in \Z^d$. From Eqs. (\ref{eq4.2}) and (\ref{eq4.3}), we get immediately
\begin{equation}
\begin{aligned}
&\lambda\Psi^{2k-1}(\xvec-\evec_k)+\Psi^{2k-1}(\xvec)=\lambda\Psi^{2k}(\xvec+\evec_k)+\Psi^{2k}(\xvec),
\end{aligned}
\label{eq4.4}
\end{equation}
for any $k=1,2,\dots,d$ and $\x \in \Z^d$. In order to state the following lemma, we introduce the support of $\Psi : \Z^d \rightarrow \C^{2d}$ as follows.

\begin{equation}
S(\Psi)=\{ \x \in \Z^d : \Psi(\x) \ne \zero \}.
\label{eq4.5}
\end{equation}

$\\$
{\bf Lemma 1} \quad {\it Let $\Psi \in W(\lambda)$ with $\lambda=\pm1$.
Suppose $\#(S(\Psi))<\infty$, where $\#(A)$ is the cardinality of a set $A$. If there exist $k \in \{1,2,\cdots,d \}$ and $\x \in \Z^d$ such that 
\begin{equation}
\begin{bmatrix}\Psi^{2k-1}(\x) \\ \Psi^{2k}(\x) \end{bmatrix} \ne \begin{bmatrix}0 \\ 0 \end{bmatrix},
\label{eq4.6}
\end{equation}
then we have
\begin{equation}
\begin{bmatrix}\Psi^{2k-1}(\x-\e_k) \\ \Psi^{2k}(\x-\e_k) \end{bmatrix} \ne \begin{bmatrix}0 \\ 0 \end{bmatrix} \  or \  
\begin{bmatrix}\Psi^{2k-1}(\x+\e_k) \\ \Psi^{2k}(\x+\e_k) \end{bmatrix} \ne \begin{bmatrix}0 \\ 0 \end{bmatrix}.
\label{eq4.7}
\end{equation}
}
$\\$
\noindent
{\bf Proof} \quad First we assume that 
\begin{equation}
\begin{bmatrix}\Psi^{2k-1}(\x) \\ \Psi^{2k}(\x) \end{bmatrix}\ne \begin{bmatrix}0 \\ 0 \end{bmatrix},
\label{eq4.8}
\end{equation}
for some $k \in \{ 1,2,\cdots ,d \}$ and $\x \in \Z^d$. Moreover we suppose
\begin{equation}
\begin{bmatrix}\Psi^{2k-1}(\x-\e_k) \\ \Psi^{2k}(\x-\e_k) \end{bmatrix}=\begin{bmatrix}0 \\ 0 \end{bmatrix}
{\rm \ and \ }
\begin{bmatrix}\Psi^{2k-1}(\x+\e_k) \\ \Psi^{2k}(\x+\e_k) \end{bmatrix}=\begin{bmatrix}0 \\ 0 \end{bmatrix},
\label{eq4.9}
\end{equation}
that is,
\begin{equation}
\Psi^{2k-1}(\x-\e_k)=0,\\
\label{eq4.10}
\end{equation}
\begin{equation}
\Psi^{2k}(\x-\e_k)=0,\\
\label{eq4.11}
\end{equation}
\begin{equation}
\Psi^{2k-1}(\x+\e_k)=0,\\
\label{eq4.12}
\end{equation}
\begin{equation}
\Psi^{2k}(\x+\e_k)=0.
\label{eq4.13}
\end{equation}
Combining Eq. \eqref{eq4.4} with Eqs. \eqref{eq4.10} and \eqref{eq4.13}, we have 
\begin{equation}
\Psi^{2k-1}(\x)=\Psi^{2k}(\x).
\label{eq4.14}
\end{equation}
From the assumption Eqs. \eqref{eq4.8} and \eqref{eq4.14}, we put
\begin{equation}
\Psi^{2k-1}(\x)=\Psi^{2k}(\x)=\eta,
\label{eq4.15}
\end{equation}
where $\eta \in \C$ with $\eta \ne 0$. Furthermore, by Eq. \eqref{eq4.4} for $\x-\e_k$, we obtain
\begin{equation}
\lambda\Psi^{2k-1}(\x-2\e_k)+\Psi^{2k-1}(\x-\e_k)=\lambda\Psi^{2k}(\x)+\Psi^{2k}(\x-\e_k).
\label{eq4.16}
\end{equation}
Combining Eq. \eqref{eq4.16} with Eqs. \eqref{eq4.10},  \eqref{eq4.11} and \eqref{eq4.15} implies
\begin{equation}
\Psi^{2k-1}(\x-2\e_k)=\eta,
\label{eq4.17}
\end{equation}
since $\lambda\ne$0. 
In a similar way, Eq. \eqref{eq4.4} for $\x-2\e_k$ becomes
\begin{equation}
\lambda\Psi^{2k-1}(\x-3\e_k)+\Psi^{2k-1}(\x-2\e_k)=\lambda\Psi^{2k}(\x-\e_k)+\Psi^{2k}(\x-2\e_k).
\label{eq4.18}
\end{equation}
From Eq. \eqref{eq4.18} with Eqs. \eqref{eq4.11} and \eqref{eq4.17}, we have
\begin{equation}
\Psi^{2k-1}(\x-3\e_k)=\lambda\{\Psi^{2k}(\x-2\e_k)-\eta\},
\label{eq4.19}
\end{equation}
since $\lambda=\pm1$. Similarly, Eq. \eqref{eq4.4} for $\x-3\e_k$ becomes
\begin{equation}
\lambda\Psi^{2k-1}(\x-4\e_k)+\Psi^{2k-1}(\x-3\e_k)=\lambda\Psi^{2k}(\x-2\e_k)+\Psi^{2k}(\x-3\e_k).
\label{eq4.20}
\end{equation}
From Eq. \eqref{eq4.20} with Eq. \eqref{eq4.19}, we get
\begin{equation}
\Psi^{2k-1}(\x-4\e_k)=\lambda\Psi^{2k}(\x-3\e_k)+\eta.
\label{eq4.21}
\end{equation}
Continuing this argument repeatedly, we finally abtain
\begin{equation}
\Psi^{2k-1}(\x-(j+1)\e_k)=\lambda\Psi^{2k}(\x-j\e_k)+(-\lambda)^{j+1}\eta,
\label{eq4.22}
\end{equation}
for any $j=0,1,2,\cdots$. Assumption $\#(S(\Psi))<\infty$ implies that there exists $J$ such that 
\begin{equation}
\Psi^{2k-1}(\x-j'\e_k)=\Psi^{2k}(\x-j'\e_k)=0,
\label{eq4.23}
\end{equation}
for any $j' \ge J$. Combining Eq. \eqref{eq4.22} with Eq. \eqref{eq4.23} gives $\eta=0$ since $\lambda \ne 0$. Therefore contradiction occurs, so the proof is complete.

$\\$ 
\noindent
{\bf Lemma 2} \quad {\it Let $\Psi \in W(\lambda)$ with $\lambda=\pm1$. Suppose $\#(S(\Psi))<\infty$.
If there exist $k \in \{1,2,\cdots,d\}$ and $\x \in \Z^d$ such that
\begin{align}
\begin{bmatrix}\Psi^{2k-1}(\xvec)\\ \Psi^{2k}(\xvec) \end{bmatrix} \ne 
\begin{bmatrix}0\\ 0\end{bmatrix},
\label {eq4.24}
\end{align}
then there exist $m^{(-)} (\le 0)$ and $m^{(+)} (\ge 0)$ with $m^{(-)} < m^{(+)}$ and $\ \alpha,\beta \in \C$ with $\alpha\beta \ne 0$ such that}

\begin{equation}
\begin{bmatrix}\Psi^{2k-1}(\x +m\e_k)\\ \Psi^{2k}(\x +m\e_k) \end{bmatrix} = 
\left\{
\begin{aligned}
&^T \begin{bmatrix}0, 0\end{bmatrix} \quad (m<m^{(-)} ) \\
&^T \begin{bmatrix}\alpha, 0\end{bmatrix} \quad (m=m^{(-)} ) \\
&^T \begin{bmatrix}0, \beta\end{bmatrix} \quad (m=m^{(+)} ) \\
&^T \begin{bmatrix}0, 0\end{bmatrix} \quad (m>m^{(+)} )
\end{aligned}
\label {eq4.25}
\right. \  .
\end{equation}
{\it Moreover, we have}
\begin{equation}
\begin{aligned}
\begin{bmatrix}\Psi^{2l-1}(\x+m^{(-)}\e_k)\\ \Psi^{2l}(\x+m^{(-)}\e_k) \end{bmatrix} \ne \begin{bmatrix}0\\ 0\end{bmatrix},
\end{aligned}
\label {eq4.26}
\end{equation}
{\it and}
\begin{equation}
\begin{aligned}
\begin{bmatrix}\Psi^{2l-1}(\x+m^{(+)}\e_k)\\ \Psi^{2l}(\x+m^{(+)}\e_k) \end{bmatrix} \ne \begin{bmatrix}0\\ 0\end{bmatrix},
\end{aligned}
\label {eq4.27}
\end{equation}
{\it for any $l \in \{1,2,\cdots,d \}\backslash \{ k \} $.}

$\\$
\noindent
{\bf Proof} \quad From Lemma 1, we get $\# S(\Psi)\ge 2$. Therefore we see that there exist $m^{(-)} (\le 0)$ and $m^{(+)} (\ge 0)$ with $m^{(-)}<m^{(+)}$ and $\alpha,\beta,\gamma,\delta \in \C$ with $|\alpha|+|\gamma|>0$ and $|\beta|+|\delta|>0$ such that

\begin{equation}
\begin{bmatrix}\Psi^{2k-1}(\x +m\e_k)\\ \Psi^{2k}(\x +m\e_k) \end{bmatrix} = 
\left\{
\begin{aligned}
&^T \begin{bmatrix}0, 0\end{bmatrix} \quad &(m<m^{(-)} ) \\
&^T \begin{bmatrix}\alpha, \gamma\end{bmatrix} \quad &(m=m^{(-)} ) \\
&^T \begin{bmatrix}\delta, \beta\end{bmatrix} \quad &(m=m^{(+)} ) \\
&^T \begin{bmatrix}0, 0\end{bmatrix} \quad &(m>m^{(+)} )
\end{aligned}
\right. \  .
\label {eq4.28}
\end{equation}
By Eq. \eqref{eq4.4} for $\x+(m^{(-)} -1)\e_k$, we have 
\begin{equation}
\begin{aligned}
\lambda\Psi^{2k-1}(\x+(m^{(-)}-2)\e_k)+\Psi^{2k-1}(\x+(m^{(-)}-1)\e_k) \\
=\lambda\Psi^{2k}(\x+m^{(-)}\e_k)+\Psi^{2k}(\x+(m^{(-)}-1)\e_k).
\label {eq4.29}
\end{aligned}
\end{equation}
\noindent
Combining Eq. \eqref{eq4.28} with Eq. \eqref{eq4.29} gives
\begin{equation}
\Psi^{2k}(\x+m^{(-)}\e_k)=\gamma=0,
\label {eq4.30}
\end{equation}
since $\lambda \ne 0$. In a similar fashion, from Eq. \eqref{eq4.4} for $\x+(m^{(+)}+1)\e_k$, we have
\begin{equation}
\begin{aligned}
&\Psi^{2k-1}(\x+m^{(+)}\e_k)=\delta=0.
\label {eq4.31}
\end{aligned}
\end{equation}
\noindent
Thus combining Eqs. \eqref{eq4.28}, \eqref{eq4.30} and \eqref{eq4.31} implies Eq. \eqref{eq4.25}.
$\\$
$\\$
\noindent

By Eq. \eqref{eq4.2} for $\x+m^{(-)}\e_k$, we have
\begin{equation}
\lambda\Psi^{2k-1}(\x+(m^{(-)}-1)\e_k)+\Psi^{2k-1}(\x+m^{(-)}\e_k)=\frac{1}{d}\Gamma(\x+m^{(-)}\e_k).
\label {eq4.32}
\end{equation}
Then combining Eq. \eqref{eq4.32} with Eq. \eqref{eq4.25} gives
\begin{equation}
\frac{1}{d}\Gamma(\x+m^{(-)}\e_k)=\alpha.
\label {eq4.33}
\end{equation}
Similarly, by Eq. \eqref{eq4.3} for $\x+m^{(+)}\e_k$ and Eq. \eqref{eq4.25}, we get
\begin{equation}
\frac{1}{d}\Gamma(\x+m^{(+)}\e_k)=\beta.
\label {eq4.34}
\end{equation}
From now on, we assume that there exists $l \in \{ 1,2,\cdots,d \}\backslash \{ k\}$ such that 
\begin{equation}
\begin{bmatrix}\Psi^{2l-1}(\x+m^{(-)}\e_k)\\
\Psi^{2l}(\x+m^{(-)}\e_k) \end{bmatrix} = \begin{bmatrix}0\\ 0\end{bmatrix},
\label {eq4.35}
\end{equation}
or 
\begin{equation}
\begin{bmatrix}\Psi^{2l-1}(\x+m^{(+)}\e_k)\\
\Psi^{2l}(\x+m^{(+)}\e_k) \end{bmatrix} = \begin{bmatrix}0\\ 0\end{bmatrix}.
\label {eq4.36}
\end{equation}
First we consider Eq. \eqref{eq4.35} case. We now use Eq. \eqref{eq4.2} with 
$k \rightarrow l$ and $\x \rightarrow \x+m^{(-)}\e_k$ to get 
\begin{equation}
\lambda\Psi^{2l-1}((\x+m^{(-)}\e_k)-\e_l)+\Psi^{2l-1}(\x+m^{(-)}\e_k)=\frac{1}{d}\Gamma(\x+m^{(-)}\e_k).
\label {eq4.37}
\end{equation}
Using the equation just derived and Eq. \eqref{eq4.33}, we have
\begin{equation}
\lambda\Psi^{2l-1}((\x+m^{(-)}\e_k)-\e_l)+\Psi^{2l-1}(\x+m^{(-)}\e_k)=\alpha.
\label {eq4.38}
\end{equation}
By assumption $\Psi^{2l-1}(\x+m^{(-)}\e_k)=0$ in Eq. \eqref{eq4.35}, we see that Eq. \eqref{eq4.38} becomes
\begin{equation}
\Psi^{2l-1}((\x+m^{(-)}\e_k)-\e_l)=\lambda\alpha,
\label {eq4.39}
\end{equation}
since $\lambda=\pm1$.
Next we see Eq. \eqref{eq4.4} with $k \rightarrow l$ and $\x \rightarrow \x+m^{(-)}\e_k-\e_l$ to get 
\begin{equation}
\begin{aligned}
\lambda\Psi^{2l-1}((\x+m^{(-)}\e_k)-2\e_l)+\Psi^{2l-1}((\x+m^{(-)}\e_k)-\e_l)\\
=\lambda\Psi^{2l}(\x+m^{(-)}\e_k)+\Psi^{2l}((\x+m^{(-)}\e_k)-\e_l).
\label {eq4.40}
\end{aligned}
\end{equation}
Combining this equation with Eq. \eqref{eq4.39} and assumption $\Psi^{2l}(\x+m^{(-)}\e_k)=0$ in Eq. \eqref{eq4.35} gives
\begin{equation}
\Psi^{2l-1}((\x+m^{(-)}\e_k)-2\e_l)=\lambda\Psi^{2l}((\x+m^{(-)}\e_k)-\e_l)-\lambda^2\alpha,
\label {eq4.41}
\end{equation}
since $\lambda=\pm1$.
By the similar argument repeatedly, we obtain,
\begin{equation}
\Psi^{2l-1}((\x+m^{(-)}\e_k)-(j+1)\e_l)=\lambda\Psi^{2l}((\x+m^{(-)}\e_k)-j\e_l)-(-\lambda)^{j+1}\alpha,
\label {eq4.42}
\end{equation}
for any $j=1,2,\cdots$. Assumption $\#(S(\Psi))<\infty$ implies that there exists $J$ such that
\begin{equation}
\Psi^{2l-1}((\x+m^{(-)}\e_k)-j'\e_k)=\Psi^{2l}((\x+m^{(-)}\e_k)-j'\e_l)=0,
\label {eq4.43}
\end{equation}
for any $j'\ge J$. Combining Eq. \eqref{eq4.42} with Eq. \eqref{eq4.43} gives $\alpha=0$ since $\lambda \ne 0$. Thus we have a contradiction.
\par

Next we consider Eq. \eqref{eq4.36} case. In a simiar fashion, we get $\beta=0$ and have a contradiction. Therefore the proof of Lemma 2 is complete.

$\\$ 
{\bf Theorem 1} {\it For the Grover walk on $\Z^d$ with moving shift, we have}
\begin{equation}
\begin{aligned}
\#(S(\Psi)) \ge 2^d,
\end{aligned}
\label {eq4.44}
\end{equation}
{\it for any $\Psi \in W(\lambda)$ with $\lambda=\pm1$. In particular, there exists $\Psi^{(\lambda)}_\star \in W(\lambda)$ such that}
\begin{equation}
\#(S(\Psi^{(\lambda)}_\star))=2^d,
\label {eq4.45}
\end{equation}
{\it for $\lambda=\pm1$. In fact, we obtain}
\begin{equation}
\Psi^{(\lambda)}_\star(\x)=\lambda^{x_1+x_2+\cdots+x_d}
\times \ ^{T}\begin{bmatrix} |x_1 \rangle, \ |x_2 \rangle, \ \cdots ,\ |x_d \rangle \end{bmatrix} 
\quad (\x \in S(\Psi^{(\lambda)}_\star)),
\label {eq4.46}
\end{equation}
{\it where}
\begin{equation}
S(\Psi^{(\lambda)}_\star)=\{ \x=(x_1,x_2,\cdots,x_d) \in \Z^{d}:x_k \in \{0,1\} \ (k = 1,2,\cdots,d)\}.
\label {eq4.47}
\end{equation}
{\it Here $\ |0 \rangle= \ ^{T}[1, \ 0]$ and $|1 \rangle= \ ^{T}[0, \ 1]$}.

$\\$ 
{\bf Proof.} \quad For $\Psi \in W(\lambda)$ with $\lambda=\pm1$, there exist $k \in \{1,2,\cdots,d\}$ and $\x \in \Z^d$ such that  
\begin{equation}
\begin{bmatrix}\Psi^{2k-1}(\x) \\ \Psi^{2k}(\x) \end{bmatrix}\ne \begin{bmatrix}0 \\ 0 \end{bmatrix}.
\label{eq4.48}
\end{equation}
Thus, we have $\x \in S(\Psi)$. 

First we consider $d=1$ case. From Lemma 1, we see that
$\x-\e_1 \in S(\Psi)$ or $\x+\e_1 \in S(\Psi)$, so $\#(S(\Psi)) \ge 2$. If fact, we can construct a $\Psi^{(\lambda)}_\star \in W(\lambda)$ with $\lambda=\pm1$ satisfying $\#(S(\Psi^{(\lambda)}_\star))=2$ as follows.

\begin{equation}
\begin{bmatrix}\Psi^{1}(\x +m_1\e_1)\\ \Psi^{2}(\x +m_1\e_1) \end{bmatrix} = 
\left\{
\begin{aligned}
&^T \begin{bmatrix}0, 0\end{bmatrix} \quad &(m_1<0) \\
&\lambda^{m_1} \times ^T \begin{bmatrix}1, 0\end{bmatrix} \quad &(m_1=0) \\
&\lambda^{m_1} \times ^T \begin{bmatrix}0, 1\end{bmatrix} \quad &(m_1=1) \\
&^T \begin{bmatrix}0, 0\end{bmatrix} \quad &(m_1>1)
\end{aligned}
\label {eq4.49}
\right. \ ,
\end{equation}
where $m_1 \in \Z$.

Next we deal with $d=2$ case. Considering the argument for $d=1$ case, we can assume
\begin{equation}
\begin{bmatrix}\Psi^{1}(\x) \\ \Psi^{2}(\x) \end{bmatrix} \ne \begin{bmatrix}0 \\ 0 \end{bmatrix} 
\  {\rm and} \   
\begin{bmatrix}\Psi^{1}(\x+\e_1) \\ \Psi^{2}(\x+\e_1) \end{bmatrix} \ne \begin{bmatrix}0 \\ 0 \end{bmatrix}.
\label{eq4.50}
\end{equation}
By Lemma 2 with Eq. \eqref{eq4.50}, we can also assume $m^{(-)}=0$ and $m^{(+)}=1$ to minimize the $\#(S(\Psi))$, then we have 
\begin{equation}
\begin{bmatrix}\Psi^{3}(\x) \\ \Psi^{4}(\x) \end{bmatrix} \ne \begin{bmatrix}0 \\ 0 \end{bmatrix} ,
\label{eq4.51}
\end{equation}
and
\begin{equation}
\begin{bmatrix}\Psi^{3}(\x+\e_1) \\ \Psi^{4}(\x+\e_1) \end{bmatrix} \ne \begin{bmatrix}0 \\ 0 \end{bmatrix}.
\label{eq4.52}
\end{equation}
From Lemma 1 with Eqs. \eqref{eq4.51} and \eqref{eq4.52}, we obtain ``$\x-\e_2 \in S(\Psi)$ or $\x+\e_2 \in S(\Psi)$'' and ``$\x+\e_1-\e_2 \in S(\Psi)$ or $\x+\e_1+\e_2 \in S(\Psi)$'' respectively, so $\#(S(\Psi))\ge4$. In fact, we can construct a $\Psi^{(\lambda)}_\star \in W(\lambda)$ with $\lambda=\pm1$ satisfying $\#(S(\Psi^{(\lambda)}_\star))=4$ as follows.
\begin{equation}
\Psi(\x +m_1\e_1+m_2\e_2)=
\left\{
\begin{aligned}
&\lambda^{m_1+m_2} \times ^T \begin{bmatrix}1, 0, 1, 0\end{bmatrix} \quad &(m_1, m_2)=(0,0) \\
&\lambda^{m_1+m_2} \times ^T \begin{bmatrix}0, 1, 1, 0\end{bmatrix} \quad &(m_1, m_2)=(1,0) \\
&\lambda^{m_1+m_2} \times ^T \begin{bmatrix}1, 0, 0, 1\end{bmatrix} \quad &(m_1, m_2)=(0,1) \\
&\lambda^{m_1+m_2} \times ^T \begin{bmatrix}0, 1, 0, 1\end{bmatrix} \quad &(m_1, m_2)=(1,1) \\
&^T \begin{bmatrix}0, 0, 0, 0\end{bmatrix} \quad &(otherwise)
\end{aligned}
\label {eq4.53}
\right. \ ,
\end{equation}
for $m_1, m_2 \in \Z$. Remark that Eq. \eqref{eq4.53} has been introduced in Stefanak et al. \cite{Stefanak1}. Continuing a similar argument for $d=3,4,\cdots,$ we have the desired conclusion. 

$\\$
\par

From Eq. \eqref{eq4.53}, we obtain the following equation as a stationary measure of Grover walk on $\Z^2$ when $\lambda=1$.
\begin{equation}
\Psi(x,y)=\begin{bmatrix}1 \\ 0 \\ 1 \\ 0 \end{bmatrix} g(x,y)
+\begin{bmatrix}0 \\ 1 \\ 1 \\ 0 \end{bmatrix} g(x-1,y)
+\begin{bmatrix}1 \\ 0 \\ 0 \\ 1 \end{bmatrix} g(x,y-1)
+\begin{bmatrix}0 \\ 1 \\ 0 \\ 1 \end{bmatrix} g(x-1,y-1) \ , 
\label {eq4.54}
\end{equation}
for $(x,y) \in \Z^2$. Here $g:\Z^2 \longrightarrow \C$. Let $g(x,y)$ as follows.
\begin{equation}
g(x,y)=
\left\{
\begin{aligned}
&\delta_{(x,y)} \quad &(x,y \in \{0,-1\}) \\
&0 \quad &(otherwise)
\end{aligned}
\right. \ . 
\label {eq4.55}
\end{equation}
Then  combining Eq. \eqref{eq4.54} with Eq. \eqref{eq4.55}, we easily get $\#(S(\Psi))=9$ 
such that
\begin{equation}
\begin{aligned}
\Psi
&=\begin{bmatrix}2 \\ 2 \\ 2 \\ 2 \end{bmatrix} \delta_{(0,0)}
+\begin{bmatrix}1 \\ 1 \\ 0 \\ 2 \end{bmatrix} \delta_{(0,1)} 
+\begin{bmatrix}0 \\ 2 \\ 1 \\ 1 \end{bmatrix} \delta_{(1,0)} 
+\begin{bmatrix}1 \\ 1 \\ 2 \\ 0 \end{bmatrix} \delta_{(0,-1)} 
+\begin{bmatrix}2 \\ 0 \\ 1 \\ 1 \end{bmatrix} \delta_{(-1,0)} \\
&+\begin{bmatrix}0 \\ 1 \\ 0 \\ 1 \end{bmatrix} \delta_{(1,1)} 
+\begin{bmatrix}0 \\ 1 \\ 1 \\ 0 \end{bmatrix} \delta_{(1,-1)} 
+\begin{bmatrix}1 \\ 0 \\ 1 \\ 0 \end{bmatrix} \delta_{(-1,-1)} 
+\begin{bmatrix}1 \\ 0 \\ 0 \\ 1 \end{bmatrix} \delta_{(-1,1)} \ .
\label {eq4.56}
\end{aligned}
\end{equation}
Remark that Eq. \eqref{eq4.56} has been introduced  in Komatsu and Konno \cite{Komatsu and Konno.}.

\section{Grover walk on $\Z^d$ with flip-flop shift}

In this section, we consider the case of the $d$-dimensional Grover walk with flip-flop shift. The eigenvalue problem $U_G\Psi=\lambda\Psi \ (\lambda \in \C$ with $|\lambda|=1)$ is equivalent to
\begin{equation}
\left\{\quad
\begin{gathered}
\lambda\Psi^{1}(\xvec)
=\frac{1}{d}\Psi^{1}(\xvec+\evec_1)+\frac{1-d}{d}\Psi^{2}(\xvec+\evec_1)
+\cdots
+\frac{1}{d}\Psi^{2d-1}(\xvec+\evec_1)+\frac{1}{d}\Psi^{2d}(\xvec+\evec_1),\\
\lambda\Psi^{2}(\xvec)
=\frac{1-d}{d}\Psi^{1}(\xvec-\evec_1)+\frac{1}{d}\Psi^{2}(\xvec-\evec_1)
+\cdots
+\frac{1}{d}\Psi^{2d-1}(\xvec-\evec_1)+\frac{1}{d}\Psi^{2d}(\xvec-\evec_1),\\
\vdots\\
\lambda\Psi^{2d-1}(\xvec)
=\frac{1}{d}\Psi^{1}(\xvec+\evec_d)+\frac{1}{d}\Psi^{2}(\xvec+\evec_d)
+\cdots
+\frac{1}{d}\Psi^{2d-1}(\xvec+\evec_d)+\frac{1-d}{d}\Psi^{2d}(\xvec+\evec_d),\\
\lambda\Psi^{2d}(\xvec)
=\frac{1}{d}\Psi^{1}(\xvec-\evec_d)+\frac{1}{d}\Psi^{2}(\xvec-\evec_d)
+\cdots
+\frac{1-d}{d}\Psi^{2d-1}(\xvec-\evec_d)+\frac{1}{d}\Psi^{2d}(\xvec-\evec_d),\\
\end{gathered}
\right.
\label{eq5.1}
\end{equation}
where $\Psi(\xvec)={}^{T} \begin{bmatrix}\Psi^{1}(\xvec), \Psi^{2}(\xvec), \cdots, \Psi^{2d}(\xvec)\end{bmatrix}$ $(\x \in \Z^d)$. Put ${\displaystyle \Gamma(\x)= \sum_{j=1}^{2d} \Psi^{j} (\x)}$ for $\x\in\Z^d$. By using $\Gamma(\x)$, Eq. \eqref{eq5.1} can be written as

\begin{equation}
\begin{aligned}
\lambda\Psi^{2k-1}(\xvec-\evec_k)+\Psi^{2k}(\xvec)=\frac{1}{d}\Gamma(\xvec),\\
\end{aligned}
\label{eq5.2}
\end{equation}
\begin{equation}
\begin{aligned}
\lambda\Psi^{2k}(\xvec+\evec_k)+\Psi^{2k-1}(\xvec)=\frac{1}{d}\Gamma(\xvec),
\end{aligned}
\label{eq5.3}
\end{equation}
for any $k=1,2,\dots,d$ and $x \in \Z^d$. From Eqs. \eqref{eq5.2} and \eqref{eq5.3}, we get immediately
\begin{equation}
\begin{aligned}
&\lambda\Psi^{2k-1}(\xvec-\evec_k)+\Psi^{2k}(\xvec)=\lambda\Psi^{2k}(\xvec+\evec_k)+\Psi^{2k-1}(\xvec),
\end{aligned}
\label{eq5.4}
\end{equation}
for any $k=1,2,\dots,d$ and $x \in \Z^d$.

$\\$
{\bf Lemma 3} \quad {\it Let $\Psi \in W(\lambda)$ with $\lambda=\pm1$. Suppose $\#(S(\Psi))<\infty$. If there exist $k \in \{1,2,\cdots,d \}$ and $\x \in \Z^d$ such that
\begin{equation}
\begin{bmatrix}\Psi^{2k-1}(\x) \\ \Psi^{2k}(\x) \end{bmatrix} \ne \begin{bmatrix}0 \\ 0 \end{bmatrix},
\label{eq5.5}
\end{equation}
then we have
\begin{equation}
\begin{bmatrix}\Psi^{2k-1}(\x-\e_k) \\ \Psi^{2k}(\x-\e_k) \end{bmatrix} \ne \begin{bmatrix}0 \\ 0 \end{bmatrix} \  or \  
\begin{bmatrix}\Psi^{2k-1}(\x+\e_k) \\ \Psi^{2k}(\x+\e_k) \end{bmatrix} \ne \begin{bmatrix}0 \\ 0 \end{bmatrix}.
\label{eq5.6}
\end{equation}
}
$\\$
\noindent
{\bf Proof} \quad First we assume that 
\begin{equation}
\begin{bmatrix}\Psi^{2k-1}(\x) \\ \Psi^{2k}(\x) \end{bmatrix}\ne \begin{bmatrix}0 \\ 0 \end{bmatrix},
\label{eq5.7}
\end{equation}
for some $k \in \{ 1,2,\cdots ,d \}$ and $\x \in \Z^d$. Furthermore we suppose
\begin{equation}
\begin{bmatrix}\Psi^{2k-1}(\x-\e_k) \\ \Psi^{2k}(\x-\e_k) \end{bmatrix}=\begin{bmatrix}0 \\ 0 \end{bmatrix}
{\rm \ and \ }
\begin{bmatrix}\Psi^{2k-1}(\x+\e_k) \\ \Psi^{2k}(\x+\e_k) \end{bmatrix}=\begin{bmatrix}0 \\ 0 \end{bmatrix}.
\label{eq5.8}
\end{equation}
By a similar calculation as in Lemma 1, we get the following equation corresponding to Eq. \eqref{eq4.22}.
\begin{equation}
\Psi^{2k-1}(\x-(j+1)\e_k)=-\lambda\Psi^{2k}(\x-j\e_k)+\lambda^{j+1}\eta,
\label{eq5.9}
\end{equation}
where $\eta=\Psi^{2k-1}(\x)=\Psi^{2k}(\x)$ for any $j=0,1,2,\cdots$. Assumption $\#(S(\Psi))<\infty$ implies that there exists $J$ such that 
\begin{equation}
\Psi^{2k-1}(\x-j'\e_k)=\Psi^{2k}(\x-j'\e_k)=0,
\label{eq5.10}
\end{equation}
for any $j' \ge J$. Combining Eq. \eqref{eq5.9} with Eq. \eqref{eq5.10} gives $\eta=0$ since $\lambda \ne 0$. Therefore contradiction occurs, so the proof is complete.

$\\$ 
\noindent
{\bf Lemma 4} \quad {\it Let $\Psi \in W(\lambda)$ with $\lambda=\pm1$. Suppose $\#(S(\Psi))<\infty$.
If there exist $k \in \{1,2,\cdots,d\}$ and $\x \in \Z^d$ such that
\begin{align}
\begin{bmatrix}\Psi^{2k-1}(\xvec)\\ \Psi^{2k}(\xvec) \end{bmatrix} \ne 
\begin{bmatrix}0\\ 0\end{bmatrix},
\label {eq5.11}
\end{align}
then there exist $m^{(-)} (\le 0)$ and $m^{(+)} (\ge 0)$ with $m^{(-)} < m^{(+)}$ and $\ \alpha,\beta \in \C$ with $\alpha\beta \ne 0$ such that}

\begin{equation}
\begin{bmatrix}\Psi^{2k-1}(\x +m\e_k)\\ \Psi^{2k}(\x +m\e_k) \end{bmatrix} = 
\left\{
\begin{aligned}
&^T \begin{bmatrix}0, 0\end{bmatrix} \quad (m<m^{(-)} ) \\
&^T \begin{bmatrix}\alpha, 0\end{bmatrix} \quad (m=m^{(-)} ) \\
&^T \begin{bmatrix}0, \beta\end{bmatrix} \quad (m=m^{(+)} ) \\
&^T \begin{bmatrix}0, 0\end{bmatrix} \quad (m>m^{(+)} )
\end{aligned}
\label {eq5.12}
\right. \  .
\end{equation}
{\it Moreover, we have}
\begin{equation}
\Gamma(\x+m^{(-)}\e_k) = 0 \ ,
\label {eq5.13}
\end{equation}
{\it and}
\begin{equation}
\Gamma(\x+m^{(+)}\e_k) = 0 \ .
\label {eq5.14}
\end{equation}

$\\$
\noindent
{\bf Proof} \quad From Lemma 3, we get $\# S(\Psi)\ge 2$. Therefore we see that there exist $m^{(-)} (\le 0)$ and $m^{(+)} (\ge 0)$ with $m^{(-)}<m^{(+)}$ and $\alpha,\beta,\gamma,\delta \in \C$ with $|\alpha|+|\gamma|>0$ and $|\beta|+|\delta|>0$ such that

\begin{equation}
\begin{bmatrix}\Psi^{2k-1}(\x +m\e_k)\\ \Psi^{2k}(\x +m\e_k) \end{bmatrix} = 
\left\{
\begin{aligned}
&^T \begin{bmatrix}0, 0\end{bmatrix} \quad &(m<m^{(-)} ) \\
&^T \begin{bmatrix}\alpha, \gamma\end{bmatrix} \quad &(m=m^{(-)} ) \\
&^T \begin{bmatrix}\delta, \beta\end{bmatrix} \quad &(m=m^{(+)} ) \\
&^T \begin{bmatrix}0, 0\end{bmatrix} \quad &(m>m^{(+)} )
\end{aligned}
\right. \  .
\label {eq5.15}
\end{equation}
By Eq. \eqref{eq5.4} for $\x+(m^{(-)} -1)\e_k$, we have 
\begin{equation}
\begin{aligned}
\lambda\Psi^{2k-1}(\x+(m^{(-)}-2)\e_k)+\Psi^{2k}(\x+(m^{(-)}-1)\e_k) \\
=\lambda\Psi^{2k}(\x+m^{(-)}\e_k)+\Psi^{2k-1}(\x+(m^{(-)}-1)\e_k).
\label {eq5.16}
\end{aligned}
\end{equation}
\noindent
Combining Eq. \eqref{eq5.15} with Eq. \eqref{eq5.16} gives
\begin{equation}
\Psi^{2k}(\x+m^{(-)}\e_k)=\gamma=0,
\label {eq5.17}
\end{equation}
since $\lambda \ne 0$. In a similar fashion, from Eq. \eqref{eq5.4} for $\x+(m^{(+)}+1)\e_k$, we have
\begin{equation}
\begin{aligned}
&\Psi^{2k-1}(\x+m^{(+)}\e_k)=\delta=0.
\label {eq5.18}
\end{aligned}
\end{equation}
\noindent
Thus combining Eqs. \eqref{eq5.15}, \eqref{eq5.17} and \eqref{eq5.18} implies Eq. \eqref{eq5.12}.
$\\$
$\\$
\noindent

By Eq. \eqref{eq5.2} for $\x+m^{(-)}\e_k$, we have
\begin{equation}
\lambda\Psi^{2k-1}(\x+(m^{(-)}-1)\e_k)+\Psi^{2k}(\x+m^{(-)}\e_k)=\frac{1}{d}\Gamma(\x+m^{(-)}\e_k).
\label {eq5.19}
\end{equation}
Then combining Eq. \eqref{eq5.19} with Eq. \eqref{eq5.12} gives
\begin{equation}
\frac{1}{d}\Gamma(\x+m^{(-)}\e_k)=0.
\label {eq5.20}
\end{equation}
Similarly, by Eq. \eqref{eq5.3} for $\x+m^{(+)}\e_k$ and Eq. \eqref{eq5.12}, we get
\begin{equation}
\frac{1}{d}\Gamma(\x+m^{(+)}\e_k)=0.
\label {eq5.21}
\end{equation}
Therefore the proof of Lemma 4 is complete.

$\\$ 
{\bf Theorem 2} {\it For the Grover walk on $\Z^d$ with flip-flop shift, we have}
\begin{equation}
\left\{
\begin{aligned}
&\#(S(\Psi)) = 0 \quad (d=1) \\
&\#(S(\Psi)) \ge 4 \quad (d \ge 2)
\end{aligned}
\label {eq5.22}
\right. ,
\end{equation}
{\it for any} $\Psi \in W(\lambda)$ {\it with } $\lambda=\pm1$. 
{\it In particular, there exists } $\Psi^{(\lambda)}_\star \in W(\lambda)$ {\it such that}
\begin{equation}
\#(S(\Psi)) =  4 \quad (d \ge 2)
\label {eq5.23}
\end{equation}
{\it for } $\lambda = \pm 1$. {\it In fact, we obtain}
\begin{equation}
\Psi^{(\lambda)}_\star(\x)=\lambda^{x_1+x_2}
\times \ ^{T}\begin{bmatrix}(-1)^{x_1+x_2} |x_1 \rangle ,\ (-1)^{x_1+x_2+1}|x_2 \rangle, \zero ,\cdots,\zero \end{bmatrix} 
\quad (\x \in S(\Psi^{(\lambda)}_\star)),
\label {eq5.24}
\end{equation}
{\it where}
\begin{equation}
S(\Psi^{(\lambda)}_\star)=\{ \x=(x_1,x_2,\cdots,x_d) \in \Z^{d}:x_1,x_2 \in \{0,1\}, \ x_3=x_4=\cdots=x_d = 0 \}.
\label {eq5.25}
\end{equation}
{\it Here $\ |0 \rangle= \ ^{T}[1, \ 0]$, $\ |1 \rangle= \ ^{T}[0, \ 1]$ and  $\ \zero=\ ^{T}[0, \ 0]$}.

$\\$ 
\noindent
{\bf Proof} \quad First, we consider $d=1$ case. For $\Psi \in W(\lambda)$ with $\lambda=\pm1$, there exists $x \in \Z$ such that
\begin{align}
\begin{bmatrix}\Psi^{1}(x)\\ \Psi^{2}(x) \end{bmatrix} \ne 
\begin{bmatrix}0\\ 0\end{bmatrix}.
\label {eq5.26}
\end{align}
From Lemma 4, we have $m_1^{(-)} (\le 0)$ and $m_1^{(+)} (\ge 0)$ with $m_1^{(-)} < m_1^{(+)}$ and $\alpha,\beta \in \C$ with $\alpha\beta \ne 0$ such that
\begin{equation}
\begin{bmatrix}\Psi^{1}(x +m_1)\\ \Psi^{2}(x +m_1) \end{bmatrix} = 
\left\{
\begin{aligned}
&^T \begin{bmatrix}0, 0\end{bmatrix} \quad (m_1<m_1^{(-)} ) \\
&^T \begin{bmatrix}\alpha, 0\end{bmatrix} \quad (m_1=m_1^{(-)} ) \\
&^T \begin{bmatrix}0, \beta\end{bmatrix} \quad (m_1=m_1^{(+)} ) \\
&^T \begin{bmatrix}0, 0\end{bmatrix} \quad (m_1>m_1^{(+)} )
\end{aligned}
\label {eq5.27}
\right. \  ,
\end{equation}
and
\begin{equation}
\left\{
\begin{aligned}
\Gamma(x+m_1^{(-)})=0 \\
\Gamma(x+m_1^{(+)})=0 
\end{aligned}
\label {eq5.28}
\right. \  .
\end{equation}
By definition of $\Gamma$ and Eq. \eqref{eq5.27}, we have
\begin{equation}
\left\{
\begin{aligned}
\Gamma(x+m_1^{(-)})=\alpha \\
\Gamma(x+m_1^{(+)})=\beta 
\end{aligned}
\label {eq5.29}
\right. \  .
\end{equation}
Combining Eq. \eqref{eq5.28} with Eq. \eqref{eq5.29}, we get $\alpha=\beta=0$. So we see that the finite support for $d$=1 does not exist.
$\\$
$\\$
\indent
Next we deal with $d=2$ case. For $\Psi \in W(\lambda)$ with $\lambda=\pm1$, we assume that there exists $\x \in \Z^2$ such that 
\begin{align}
\begin{bmatrix}\Psi^{1}(\x)\\ \Psi^{2}(\x) \end{bmatrix} \ne 
\begin{bmatrix}0\\ 0\end{bmatrix},
\label {eq5.30}
\end{align}
and we put 
\begin{equation}
\left\{
\begin{aligned}
m^{(-)}=0 \\
m^{(+)}=1
\end{aligned}
\right. \  ,
\label {eq5.31}
\end{equation}
for Eq. \eqref{eq5.12} on Lemma 4 to minimize $\#(S(\Psi))$.
By using \eqref{eq5.12} with Eq. \eqref{eq5.31}, we have
\begin{align}
\begin{bmatrix}\Psi^{1}(\x)\\ \Psi^{2}(\x) \end{bmatrix} =
\begin{bmatrix}\alpha \\ 0\end{bmatrix}.
\label {eq5.32}
\end{align}
By definition of $\Gamma$ with Eqs. \eqref{eq5.13}, \eqref{eq5.31} and \eqref{eq5.32}, we get
\begin{align}
\begin{bmatrix}\Psi^{3}(\x)\\ \Psi^{4}(\x) \end{bmatrix} \ne 
\begin{bmatrix}0\\ 0\end{bmatrix},
\label {eq5.33}
\end{align}
since $\alpha \ne 0$.

Similarly, from Lemma 3 with Eq. \eqref{eq5.30}, we can assume
\begin{align}
\begin{bmatrix}\Psi^{1}(\x+\e_1)\\ \Psi^{2}(\x+\e_1) \end{bmatrix} \ne 
\begin{bmatrix}0\\ 0\end{bmatrix},
\label {eq5.34}
\end{align}
and we obtain
\begin{align}
\begin{bmatrix}\Psi^{3}(\x+\e_1)\\ \Psi^{4}(\x+\e_1) \end{bmatrix} \ne 
\begin{bmatrix}0\\ 0\end{bmatrix},
\label {eq5.35}
\end{align}
since $\beta \ne 0$. From Lemma 3 with Eqs. \eqref{eq5.33} and \eqref{eq5.35}, we obtain ``$\x-\e_2 \in S(\Psi)$ or $\x+\e_2 \in S(\Psi)$'' and ``$\x+\e_1-\e_2 \in S(\Psi)$ or $\x+\e_1+\e_2 \in S(\Psi)$'' respectively, so $\#(S(\Psi)) \ge 4$. In fact, we can construct a $\Psi^{(\lambda)}_\star \in W(\lambda)$ with $\lambda=\pm1$ satisfying $\#(S(\Psi^{(\lambda)}_\star))=4$ as follows.
\begin{equation}	
\Psi(\x +m_1\e_1+m_2\e_2)=
\left\{
\begin{aligned}
& ^T \begin{bmatrix}1, 0, -1, 0\end{bmatrix} \quad &(m_1, m_2)=(0,0) \\
& ^T \begin{bmatrix}0, -\lambda, \lambda, 0\end{bmatrix} \quad &(m_1, m_2)=(1,0) \\
& ^T \begin{bmatrix}-\lambda, 0, 0, \lambda\end{bmatrix} \quad &(m_1, m_2)=(0,1) \\
& ^T \begin{bmatrix}0, 1, 0, -1\end{bmatrix} \quad &(m_1, m_2)=(1,1) \\
& ^T \begin{bmatrix}0, 0, 0, 0\end{bmatrix} \quad &(otherwise)
\end{aligned}
\label {eq5.36}
\right. \ ,
\end{equation}
for $m_1, m_2 \in \Z$.

Finally, we consider $d\ge3$ case by continuing the argument on $d=2$ case. To expand Eq. \eqref{eq5.36} to $d\ge3$, we focus on the fact that $\Gamma(\x+m_1\e_1+m_2\e_2)=0$ for any $\x \in \Z^2$ and $m_1,m_2 \in \Z$ in Eq. \eqref{eq5.36}. By assuming $\Psi^{2k-1}(\x+m_1\e_1+m_2\e_2)=\Psi^{2k}(\x+m_1\e_1+m_2\e_2)=0$ for any $k \in \{3,4,\cdots,d\}$, we can contruct a $\Psi^{(\lambda)}_\star \in W(\lambda)$ with $\lambda=\pm1$ satisfying $\#(S(\Psi^{(\lambda)}_\star))=4$ as follows. 

\begin{equation}
\Psi(\x +m_1\e_1+m_2\e_2)=
\left\{
\begin{aligned}
& ^T \begin{bmatrix}1, 0, -1, 0, 0, \cdots, 0\end{bmatrix} \quad &(m_1, m_2)=(0,0) \\
& ^T \begin{bmatrix}0, -\lambda, \lambda, 0, 0, \cdots, 0\end{bmatrix} \quad &(m_1, m_2)=(1,0) \\
& ^T \begin{bmatrix}-\lambda, 0, 0, \lambda, 0, \cdots, 0\end{bmatrix} \quad &(m_1, m_2)=(0,1) \\
& ^T \begin{bmatrix}0, 1, 0, -1, 0, \cdots, 0\end{bmatrix} \quad &(m_1, m_2)=(1,1) \\
& ^T \begin{bmatrix}0, 0, 0, 0, 0, \cdots, 0\end{bmatrix} \quad &(otherwise)
\end{aligned}
\label {eq5.37}
\right. \ .
\end{equation}
Theorem 2 can be derived from another approach based on the spectral mapping theorem, see Corollary 2 in Higuchi et al. \cite{Higuchi et al.}.
$
\\
$

\section{Summary}
We presented the minimum supports of states for the Grover walk on $\Z^d$ with moving and flip-flop shifts, respectively, by solving the eigenvalue problem $U_G\Psi=\lambda\Psi$. Results on the moving shift model was obtained by Theorem 1 which coincides with result in Stefanak et al. \cite{Stefanak1} ($\Z^2$ case) and improves result in Komatsu and Konno \cite{Komatsu and Konno.} ($\Z^d$ case). Moreover, results on the flip-flop shift model shown by Higuchi et al. \cite{Higuchi et al.} was given by Theorem 2. One of the interesting future problems might be to clarify a relationship between the stationary measure and the time-averaged limit measure of the Grover walk on $\Z^d$.


$\\$


\begin{thebibliography}{9}

\bibitem{Aharonov1}
Aharonov, D., Ambainis, A., Kempe, J., Vazirani, U. V., Quantum walks on graphs, Proceedings of ACM Symposium on Theory of Computation (STOC'01), July 2001, pp.50-59 (2001)

\bibitem{Higuchi et al.}
Higuchi, Y., Konno, N., Sato, I., Segawa, E., Spectral and asymptotic properties of Grover walks on crystal lattice,\ Journal of Functional Analysis,\ {\bf267},  4197-4235\ (2014)

\bibitem{Inui et al.}
Inui, N., Konishi, Y., Konno, N.,  Localization of two-dimensional quantum walks, \ Physical Review A,\ {\bf69}, 052323\ (2004)

\bibitem{Komatsu and Konno.}
Komatsu, T., Konno, N., Stationary amplitudes of quantum walks on the higher-dimensional integer lattice,
Quantum Information Processing, {\bf16}, 291 (2017)

\bibitem{Komatsu and Tate.}
Komatsu, T., Tate, T., Eigenvalues of quantum walks of Grover and Fourier types, Journal of Fourier Analysis and Applications, {\bf25}, 1293-1318 (2019)

\bibitem{Konno}
Konno, N., Quantum Walks, Lecture Notes in Mathematics, {\bf1954}, 309-452, Springer (2008)

\bibitem{Stefanak1}
Stefanak, M., Kollar, B., Kiss, T., Jex I., Full revivals in 2D quantum walks, Physica Scripta, {\bf2010}, T140 (2010)

\bibitem{Watabe et al.} 
Watabe, K., Kobayashi, N., Katori, M., Konno, N., Limit distributions of two-dimensional quantum walks,\ Physical Review A,\ {\bf77}, 062331\ (2008)










%
%
%
%
%

\end{thebibliography}
\end{document}